\documentclass [12pt]{article}
\begin{document}

\title{\textbf{The trion as an exciton interacting \\ with a
carrier}}
\author{M. Combescot and O. Betbeder-Matibet
 \\ \small{\textit{GPS, Universit\'e Denis Diderot
and Universit\'e Pierre et Marie Curie,
CNRS,}}\\ \small{\textit{Tour 23, 2 place Jussieu, 75251
Paris Cedex 05, France}}}
\date{}
\maketitle

\begin{abstract}

The X$^-$ trion is essentially an electron bound to an exciton.
However, due to the composite nature of the exciton, there is no
way to write an exciton-electron interaction potential. We can
overcome this difficulty by using a commutation technique
similar to the one we introduced for excitons interacting with
excitons, which allows to take exactly into account the
close-to-boson character of the excitons. From it, we can obtain
the X$^-$ trion creation operator in terms of exciton and
electron. We can also derive the X$^-$ trion ladder diagram
between an exciton and an electron. These are the basic tools
for future works on many-body effects involving trions.
\end{abstract}

\vspace{2cm}


\newpage

The stability of semiconductor trions has been predicted long
ago [1,2]. However, their binding energies being extremely
small in bulk materials, clear experimental evidences [3-5]
of these exciton-electron bound states have been achieved
recently only, due to the development of good semiconductor
quantum wells, the reduction of dimensionality enhancing all
binding energies.

It is now possible to study these exciton-electron bound
states as well as their interactions [6-8] with other
carriers. However many-body effects with trions are even more
subtle than many-body effects with excitons: Being made of
indistinguishable carriers, the interchange of these fermions
with other carriers is quite tricky to handle properly.

We have recently developed a novel procedure [9] to treat
many-body effects between excitons without the help of any
bosonization procedure. It allows to take
\emph{exactly} into account the possible exchanges between their
components responsible for their close-to-boson character. In
this communication, we first derive a similar procedure for an
exciton interacting with electrons. It generates an
exciton-electron coupling induced by direct Coulomb processes
and an exciton-electron coupling induced by the possible
exchange of the exciton electron with the other electrons, the
interplay between the two giving rise to very many subtle
exchange Coulomb processes.

We then use this commutation technique to get the trion
creation operators in terms of (exact-)exciton and electron
and we show how it is possible to hide all tricky
exchange couplings in the prefactors of
the trion operators, provided that we force them to have a very
specific invariance.

In a last part, we derive the trion ladder diagram between
an exciton and an electron. Although it ends by being
conceptually similar to the exciton ladder diagrams [10]
between an electron and a hole, its derivation faces at first a
major difficulty due to the composite
nature of the exciton. Indeed, the electrons being
indistinguisable, there is no way to identify the
exciton-electron \emph{potential} responsible for the trion
binding, similar to the Coulomb attracting potential which
exists between electron and hole. In addition, the spins are
known to be unimportant for excitons -- if we neglect
electron-hole exchange -- while they are crucial for trions, the
singlet and triplet states having different energy. Since the
bare Coulomb scattering is spin independent, it is not clear at
first how the total electronic spin affects these trion ladder
diagrams.

Although we here speak in terms of exciton and electron, the
extension of the present work to X$^+$ trions with electron
replaced by hole is formally straightforward, although quite
heavy due to the valence band degeneracy -- and its resulting
non diagonal heavy hole-light hole Coulomb interaction.

We wish to stress that the purpose of this work is not to find a
new clever way to estimate the trion binding energies. The wide
literature on H$^-$ or H$_2^+$ in atomic physics [11] already
provides much more accuracy than needed in Solid State Physics.
What we want to do here, is to put the trion into the general
framework of an (exact-)exciton interacting with
indistinguishable carriers, in order to provide convenient tools
for further works on many-body effects involving trions.

\section{Commutation technique for an exciton
interacting with electrons}

Following the commutation technique we developed for excitons
interacting with excitons, in the case of excitons
interacting with electrons, we are led to introduce the 
two parameters
$\Xi_{n'\mathbf{k'}n\mathbf{k}}^{\mathrm{dir}}$ and
$\Lambda_{n'\mathbf{k'}n\mathbf{k}}$ defined as
\begin{equation}
\left[V_{n,\sigma_n}^\dag\,
,a_{\mathbf{k},s}^\dag
\right]=
\sum_{n',\mathbf{k}'}\Xi_{n'\mathbf{k}'n
\mathbf{k}}
^{\mathrm{dir}}\, 
B_{n',\sigma_n}^\dag\,a_{\mathbf{k'},s}^\dag\
,
\end{equation}
\begin{equation}
\left[D_{n',\sigma_{n'};n,\sigma_n}\, ,
a_{\mathbf{k},s}^\dag\right]=\delta_{\sigma_{n'},s}\,
\sum_{\mathbf{k}'}
\Lambda_{n'\mathbf{k}'n\mathbf{k}}
\,a_{\mathbf{k'},\sigma_n}^\dag\ .
\end{equation}
The ``creation potential'' $V_{n,\sigma_n}^\dag$ and the
deviation-from-boson operator
$D_{n',\sigma_{n'};n,\sigma_n}$ are the ones defined as for
excitons interacting with excitons [9], namely
$\left[H,B_{n,\sigma_n}^\dag\right]=E_n
B_{n,\sigma_n}^\dag+V_{n,\sigma_n}^\dag$ and $\left[B_{n',
\sigma_{n'}},B_{n,\sigma_n}^\dag\right]=\delta_{n,n'}\,
\delta_{\sigma_n,\sigma_{n'}}-D_{n',\sigma_{n'};n,\sigma_n}$.
$H$ is the exact semiconductor Hamiltonian and
$B_{n,\sigma_n}^\dag$ is the creation operator of the  exciton
$n=(\nu_n,\mathbf{Q}_n)$, with energy
$E_n=\epsilon_{
\nu_n}+\hbar^2\mathbf{Q}_n^2/2(m_e+m_h)$ and
electron spin
$\sigma_n$. (The hole ``spin'' being unimportant for trions, we
drop it to simplify the notations). $B_{n,\sigma_n}^\dag$ is
linked to the electron and hole creation operators
$a_{\mathbf{k},s} ^\dag$ and $b_\mathbf{k}^\dag$ by 
\begin{equation}
B_{n,\sigma_n}^\dag=\sum_{\mathbf{p}}\langle
\mathbf{p}|\varphi_{\nu_n}\rangle\,
b_{-\mathbf{p}+\alpha_h\mathbf{Q}_n}^\dag\,
a_{\mathbf{p}+\alpha_e\mathbf{Q}_n,\sigma_n}^\dag\ ,
\end{equation}
\begin{equation}
b_{\mathbf{k}_h}^\dag\,a_{\mathbf{k}_e,s}^\dag=\sum_\nu
\langle
\varphi_\nu|\alpha_h\mathbf{k}_e-\alpha_e\mathbf{k}_h\rangle\,
B_{\nu,\mathbf{k}_e+\mathbf{k}_h,s}^\dag\ ,
\end{equation}
where $\alpha_e=1-\alpha_h=m_e/(m_e+m_h)$ and
$\langle\mathbf{k}|\varphi_\nu\rangle$ is the exciton relative
motion wave function in $\mathbf{k}$ space. By using the explicit
expression of
$V_{n,\sigma_n}^\dag$ deduced from its definition
[9], eq.\ (1) leads to
\begin{equation}
\Xi_{n'\mathbf{k}'n\mathbf{k}}^{\mathrm{dir}}
=\delta_{\mathbf{Q}_{n'}+\mathbf{k}',\mathbf{Q}_n
+\mathbf{k}}\,W_{\nu_{n'}\nu_n}(\mathbf{Q}_{n'}-\mathbf{Q}_n)
\ ,
\end{equation}
\begin{equation}
W_{\nu'\nu}(\mathbf{q})=V_{\mathbf{q}}\,\langle \varphi_{\nu'}
|e^{i\alpha_h\mathbf{q}.\mathbf{r}}-e^{-i
\alpha_e\mathbf{q}.\mathbf{r}}|\varphi_\nu\rangle\ ,
\end{equation}
$V_{\mathbf{q}}$ being the Fourier transform of the
Coulomb potential. $W_{\nu'\nu}(\mathbf{q})$ characterizes the
scattering of a $\nu$ exciton into a $\nu'$ state under a
$\mathbf{q}$ Coulomb excitation. For physical understanding, let
us mention that $\Xi_{n'\mathbf{k}'n\mathbf{k}}^{\mathrm{dir}}$
also reads
\begin{equation}
\Xi_{n'\mathbf{k}'n\mathbf{k}}^{\mathrm{dir}}=
\int
d\mathbf{r}_e\,d\mathbf{r}_{e'}\,d\mathbf{r}_h
\phi_{n'}^\ast(\mathbf{r}_e,\mathbf{r}_h)\,
f_{\mathbf{k'}}^\ast(\mathbf{r}_{e'})\,\left[
\frac{e^2}{|\mathbf{r}_{e'}
-\mathbf{r}_e|}-\frac{e^2}{|\mathbf{r}_{e'}-\mathbf{r}_h|}\right]\,
\phi_n(\mathbf{r}_e,\mathbf{r}_h)\,
f_{\mathbf{k}}(\mathbf{r}_{e'})\ ,
\end{equation}
where
$f_{\mathbf{k}}(\mathbf{r})=\langle\mathbf{r}|\mathbf{k}
\rangle=e^{i\mathbf{k}.
\mathbf{r}}/\sqrt{\mathcal{V}}$ is the
free-particle wave function
while \linebreak
$\phi_n(\mathbf{r}_e,\mathbf{r}_h)=\langle\mathbf{r}_{eh}
|\varphi_{\nu_n}\rangle\,\langle\mathbf{R}_{eh}|\mathbf{Q}_n\rangle$,
with
$\mathbf{r}_{eh}=\mathbf{r}_e-\mathbf{r}_h$ and $\mathbf{R}_{eh}
=(m_e\mathbf{r}_e+m_h\mathbf{r}_h)/(m_e+m_h)$, is 
the total wave function of the $n$ exciton :
 $\Xi_{n'\mathbf{k}'n\mathbf{k}}^{\mathrm{dir}}$ thus
corresponds to the direct scattering of a ($n$,
$\mathbf{k}$) exciton-electron state into a
($n'$, $\mathbf{k}'$) state
induced by the Coulomb interactions \emph{between} the exciton
and the electron, when the $n$ and $n'$ excitons
are made with the \emph{same} electron-hole
pair $(e,h)$.

If we turn to
$\Lambda_{n'\mathbf{k}'n\mathbf{k}}$, eqs.\ (2,3) and the
explicit expression of $D_{n',\sigma';n,\sigma}$ deduced from its
definition [9] lead to 
\begin{equation}
\Lambda_{n'\mathbf{k}'n\mathbf{k}}=\delta_{\mathbf{K'},
\mathbf{K}}\,L_{\nu_{n'},\mathbf{p'};\nu_n,\mathbf{p}}\ ,
\end{equation}
\begin{equation}
L_{\nu'\mathbf{p'}\nu\mathbf{p}}
=\langle \varphi_{\nu'}|\mathbf{p}+
\alpha_e\mathbf{p'}\rangle\,\langle\mathbf{p'}+\alpha_e
\mathbf{p}|\varphi_\nu\rangle\ ,
\end{equation}
with
$\mathbf{k}=\mathbf{p}+\beta_e\mathbf{K}$,
$\mathbf{Q}_n=-\mathbf{p}+\beta_x\mathbf{K}$
and $\beta_e=1-\beta_x=m_e/(2m_e+m_h)$. We can mention that
$\Lambda_{n'\mathbf{k}'n\mathbf{k}}$ also reads
\begin{equation}
\Lambda_{n'\mathbf{k}'n\mathbf{k}}=\int
d\mathbf{r}_e\,d\mathbf{r}_{e'}\, d\mathbf{r}_h\,\phi_{n'}^\ast
(\mathbf{r}_e,\mathbf{r}_h)\,
f_{\mathbf{k'}}^\ast(\mathbf{r}_{e'})\,
\phi_n(\mathbf{r}_{e'},\mathbf{r}_h)\,
f_{\mathbf{k}}(\mathbf{r}_e)\ ,
\end{equation}
which clearly shows that the
$(n,\mathbf{k})$ and $(n',\mathbf{k}')$ states are coupled
by $\Lambda_{n'\mathbf{k'}n\mathbf{k}}$ due to their
exchange of electrons, independently from any Coulomb process.
This possible exchange of electrons also leads to
\begin{equation}
B_{n,\sigma}^\dag\,a_{\mathbf{k},s}^\dag=
-\sum_{n',\mathbf{k}'}
\Lambda_{n'\mathbf{k}'n\mathbf{k}}\,
B_{n',s}^\dag\,a_{\mathbf{k'},\sigma}^\dag
\ ,
\end{equation}
while two exchanges reduce to identity :
\begin{equation}
\sum_{n'',\mathbf{k''}}\Lambda_{n'\mathbf{k'}
n''\mathbf{k''}}\,\Lambda_{n''\mathbf{k''}
n\mathbf{k}}=\delta_{nn'}\,\delta_{\mathbf{kk'}}\
.
\end{equation}

\section{Trion creation operators}

The $\mathrm{X}^-$ trions being made of two electrons and one
hole, their creation operators \emph{a priori} write in terms
of
$b_{\mathbf{k}_h}^\dag a_{\mathbf{k}_e}^\dag a_{\mathbf{k}
_{e'}}^\dag$. According to eq.\ (4), they can also be written in
terms of $B_{n}^\dag a_{\mathbf{k}}^\dag$, with
$\mathbf{Q}_n +\mathbf{k}$ being equal to
the trion total momentum $\mathbf{K}_i$. 

Let us look for these trion creation operators as
\begin{equation}
\mathrm{T}_{i;S,S_z}^\dag=\sum_{\nu,\mathbf{p}}f_{\nu,
\mathbf{p}}^{(\eta_i;S)}\,\mathcal{T}_{\nu,\mathbf{p},\mathbf{K}_i;S,S_z}
^\dag\ ,
\end{equation}
where $i$ stands for $(\eta_i,\mathbf{K}_i)$, and the
$\mathcal{T}^\dag$'s are the 
exciton-electron creation operators defined
by
\begin{eqnarray}
\mathcal{T}_{\nu,\mathbf{p},\mathbf{K};1,\pm1}^\dag &=
&B_{\nu,-\mathbf{p}+\beta_x\mathbf{K},\pm}^\dag\,a_
{\mathbf{p}+\beta_e\mathbf{K},\pm}^\dag\ ,\nonumber
\\
\mathcal{T}
_{\nu,\mathbf{p},\mathbf{K};S,0}^\dag 
&=&
\left(B_{\nu,-\mathbf{p}+\beta_x\mathbf{K},+}^\dag\,
a_{\mathbf{p}+\beta_e\mathbf{K},-}^\dag-(-1)^S
B_{\nu,-\mathbf{p}+\beta_x\mathbf{K},-}^\dag\,
a_{\mathbf{p}+\beta_e\mathbf{K},+}
^\dag\right)/\sqrt{2}\ ,
\end{eqnarray}
with $S=(0,1)$.
Due to eqs.\ (8,11), these operators are such that
\begin{equation}
\mathcal{T}_{\nu,\mathbf{p},\mathbf{K};S,S_z}^\dag
=(-1)^S\sum_{\nu',
\mathbf{p'}}L_{\nu'\mathbf{p'}\nu\mathbf{p}}
\mathcal{T}_{\nu',\mathbf{p'},\mathbf{K};S,S_z}^\dag\ ,
\end{equation}
while, due to the indistinguishability of the electrons, the
corresponding exciton-electron states do not form an orthogonal
basis, since
\begin{equation}
\langle
v|\mathcal{T}_{\nu',\mathbf{p'},\mathbf{K'};S',S'_z}\,\mathcal{T}
_{\nu,\mathbf{p},\mathbf{K};S,S_z}^\dag
|v\rangle=\delta_{\mathbf{K'},\mathbf{K}}\,\delta_{S',S}\,
\delta_{S'_z,S_z}\left(\delta_{\nu',\nu}\,\delta_{\mathbf{p'},
\mathbf{p}}
+(-1)^S L_{\nu'\mathbf{p'}\nu\mathbf{p}}\right)\ ,
\end{equation}

We can use eq.\ (15) to replace the $f$ prefactors
in eq.\ (13) by $F$ defined as
\begin{equation}
F_{\nu,\mathbf{p}}^{(\eta_i;S)}=\frac{1}{2}
\left(f_{\nu,\mathbf{p}}^{(\eta_i;S)}
+(-1)^S\sum_{\nu',
\mathbf{p'}}L_{\nu\mathbf{p}\nu'\mathbf{p'}}
f_{\nu',\mathbf{p'}}^{(\eta_i;S)}\right)\ ,
\end{equation}
so that, due to eq.\ (12), these prefactors now verify
\begin{equation}
F_{\nu,\mathbf{p}}^{(\eta_i;S)}=(-1)^S\sum_{\nu',
\mathbf{p'}}
L_{\nu\mathbf{p}\nu'\mathbf{p'}}\,
F_{\nu',\mathbf{p'}}^{(\eta_i;S)}\ ,
\end{equation}
which just states that they stay invariant under the possible
electron exchange corresponding to eq.\ (11). From eq.\ (13),
with
$f$ replaced by $F$, and eqs.\ (16,18), one can
easily check that
\begin{equation}
\langle v|\mathcal{T}_{\nu,\mathbf{p},\mathbf{K};S,S_z}\,
\mathrm{T}_{\eta_i,\mathbf{K}_i;S_i,S_{iz}}^\dag|v\rangle=2\,
\delta_{S,S_i}\,\delta_{S_z,S_{iz}}\,\delta_{\mathbf{K},
\mathbf{K}_i}\,F_{\nu,\mathbf{p}}^{(\eta_i;S)}\ .
\end{equation}

If we now enforce $\mathrm{T}_{i;S,S_z}^\dag|v\rangle$ to be
a trion, \emph{i}.\ \emph{e}., an eigenstate of
$H$, we must have
\begin{equation}
\langle
v|\mathcal{T}_{\nu,\mathbf{p},\mathbf{K}_i;S,S_z}\,(H-\mathcal{E}_{i;S})\,
\mathrm{T}_{i;S,S_z}^\dag|v\rangle=0\ .
\end{equation}
By using eqs.\ (14,1,5,19), we then derive the ``Schr\"{o}dinger
equation'' fulfilled by the trion prefactors,
\begin{equation}
(\epsilon_\nu+\frac{\hbar^2\mathbf{p}^2}{2\mu_t}-\varepsilon
_{\eta_i;S})
F_{\nu,\mathbf{p}}^{(\eta_i;S)}+\sum_{\nu',\mathbf{p'}}
W_{\nu\nu'}(-\mathbf{p}+\mathbf{p'})
F_{\nu',\mathbf{p'}}^{(\eta_i;S)}=0\ ,
\end{equation}
where we have set $\mathcal{E}_{i;S}=\varepsilon_{\eta_i;S}
+\hbar^2\mathbf{K}_i^2/2(2m_e+m_h)$,
$\mu_t$ being the exciton-electron relative motion mass, 
$\mu_t^{-1}=m_e^{-1}+(m_e+m_h)^{-1}$.

It can be surprising to have only 
\emph{direct} Coulomb scatterings, through
$W_{\nu\nu'}(-\mathbf{p}+\mathbf{p'})$, appearing in the trion
Schr\"{o}dinger equation (21).
Exchange processes between electrons,
through $L_{\nu\mathbf{p}
\nu'\mathbf{p'}}$, seem absent. They
are actually hidden in the $F$'s, more precisely in
their invariance relation (18).

It is possible to relate the prefactors $F$ of the trion
operator to the trion orbital wave function in a quite easy way:
Indeed eqs.\ (13-14) lead to write this wave function as
\begin{eqnarray}
\Psi_{i;S}(\mathbf{r}_e,\mathbf{r}_{e'},\mathbf{r}_h)&=&
\langle\mathbf{R}_{ee'h}|\mathbf{K}_i\rangle \left[
\psi^{(\eta_i;S)}(\mathbf{r}_{eh},\mathbf{u}_{e';eh})
+(-1)^S(e\leftrightarrow e')
\right]/2
\nonumber \\ &=&
\langle\mathbf{R}_{ee'h}|\mathbf{K}_i\rangle
\,\psi^{(\eta_i;S)}(\mathbf{r}_{eh},\mathbf{u}_{e';eh})\ ,
\end{eqnarray}
where we have set
\begin{equation}
\psi^{\eta;S)}(\mathbf{r},\mathbf{u})=\sqrt{2}\sum_{\nu,
\mathbf{p}}
F^{\eta;S)}_{\nu,\mathbf{p}}\,\langle\mathbf{r}|\varphi_\nu
\rangle\,\langle\mathbf{u}|\mathbf{p}\rangle\ .
\end{equation}
$\mathbf{R}_{ee'h}=(m_e\mathbf{r}_e+m_{e'}\mathbf{r}_{e'}+m_h
\mathbf{r}_h)/(2m_e+m_h)$ is the trion center of mass
position, and
$\mathbf{u}_{e';eh}=\mathbf{r}_{e'}-\mathbf{R}_{eh}$ is the
distance between the electron $e'$ and the center of mass of the
exciton made with $(e,h)$. (The two terms of the first line of
eq.\ (22) are indeed equal due to eqs.\ (18,9)). 
Consequently, the prefactors of the trion expansion on the
exciton-electron operators are given by
\begin{equation}
\sqrt{2}\,F^{\eta;S)}_{\nu,\mathbf{p}}=\psi^{\eta;S)}_
{\nu,\mathbf{p}}=\int
d\mathbf{r}\,d\mathbf{u}\,
\langle
\varphi_\nu|\mathbf{r}\rangle\,\langle\mathbf{p}|\mathbf{u}
\rangle\,\psi^{(\eta;S)}(\mathbf{r},\mathbf{u})\ .
\end{equation}
$\psi^{(\eta;S)}_{\nu,\mathbf{p}}$ is thus the generalized
Fourier transform ``in the exciton sense'' of the relative
motion wave function $\psi^{(\eta;S)}(\mathbf{r},\mathbf{u})$.
In the standard Fourier transform, $\langle
\varphi_\nu|\mathbf{r}\rangle$ would be replaced by
$\langle\mathbf{p'} |\mathbf{r}\rangle$.

\section{Trion ladder diagram}

It is widely known [10] that excitons correspond to ``ladder
diagrams'' between one electron and one hole propagators. By
writing the semiconductor Hamiltonian as $H=H_0+V$, these
diagrams simply result from the iteration of
$(a-H)^{-1}=(a-H_0)^{-1}+(a-H)^{-1}\,V\,(a-H_0)^{-1}$
acting on one free electron-hole pair.

For trions, the problem is much more tricky. The most na\"{\i}ve
idea is to look for the diagrammatic expansion of the trion in
terms of two electron and one hole propagators, with all
possible electron-electron and electron-hole interactions
between them. In one of our previous works on interacting
excitons, namely an exciton interacting with a distant metal
[10,12,13], we have however shown that these usual free electron
and hole diagrams turn out to be extremely complicated when
compared to the ``exciton diagrams'', in which appear exciton
propagators instead of free electron and free hole propagators.
This exciton diagram procedure however faces a major difficulty
at first, since, due to the composite nature of the exciton,
there is no way to write the exact semiconductor Hamiltonian
$H$ as
$H_0'+V'$, with
$V'$ being an exciton-electron potential, so that there is no
simple iteration of $(a-H)^{-1}$ in terms of a $V'$
interaction.

It is however possible to overcome this difficulty by using
our ``commutation technique''. From the definition of
the creation potential $V_{n,\sigma_n}^\dag$, it is easy to
show that 
\begin{equation}
\frac{1}{a-H}\,B_{n,\sigma_n}^\dag=B_{n,\sigma_n}^\dag
\,\frac{1}{a-H-E_n}+\frac{1}{a-H}\,
V_{n,\sigma_n}^\dag\,\frac{1}{a-H-E_n}\ .
\end{equation}
With $a=\Omega+i\eta$, this equation, along with
eqs.\ (1,5), gives
$(\Omega-H+i\eta)^{-1}$ acting on one exciton-electron pair
as
\begin{eqnarray}
\frac{1}{\Omega-H+i\eta}\,\mathcal{T}_{\nu,\mathbf{p},\mathbf{K};
S,S_z}^\dag|v\rangle=\frac{1}{\Omega-E_{\nu,\mathbf{p},\mathbf{K}}+i\eta}
\,\left[\mathcal{T}_{\nu,\mathbf{p},\mathbf{K};S,S_z}^\dag|v
\rangle\right.\hspace{3cm}\nonumber
\\ \left.+\sum_{\nu',\mathbf{p'}}W_{\nu'\nu}(-\mathbf{p'}+
\mathbf{p})\,\frac{1}{\Omega-H+i\eta}\,
\mathcal{T}_{\nu',\mathbf{p'},\mathbf{K};S,S_z}
^\dag|v\rangle\right]\ ,
\end{eqnarray}
where
$E_{\nu,\mathbf{p},\mathbf{K}}=\epsilon_\nu+\hbar^2\mathbf{p}^2
/2\mu_t+\hbar^2\mathbf{K}^2/2(2m_e+m_h)$ is the energy of the
free exciton-electron pair.

The iteration of the above equation leads to
\begin{equation}
\frac{1}{\Omega-H+i\eta}\,\mathcal{T}_{\nu,\mathbf{p},\mathbf{K}
;S,S_z}^\dag| v\rangle=\sum_{\nu',\mathbf{p'}}A_{\nu'\mathbf{p'}
\nu\mathbf{p}}(\Omega,\mathbf{K})\,\mathcal{T}_{\nu',\mathbf{p'},
\mathbf{K};S,S_z}^\dag|v\rangle\ ,
\end{equation}
where $A_{\nu'\mathbf{p'}\nu\mathbf{p}}(\Omega,\mathbf{K})$ 
given by
\begin{eqnarray}
A_{\nu'\mathbf{p'}\nu\mathbf{p}}(\Omega,\mathbf{K})=\left[
\delta_{\nu',\nu}\,\delta_{\mathbf{p'},\mathbf{p}}+\frac{1}
{\Omega-E_{\nu',\mathbf{p'},\mathbf{K}}+i\eta}\left\{
W_{\nu'\nu}(-\mathbf{p'}+\mathbf{p})\right.\right.\hspace{4cm}
\nonumber
\\
\left.\left.+\sum_{\nu_1,\mathbf{p}_1}
\frac{W_{\nu'\nu_1}(-\mathbf{p'}+\mathbf{p}_1)\,W_{\nu_1\nu}
(-\mathbf{p}_1+\mathbf{p})}{\Omega-E_{\nu_1,\mathbf{p}_1,
\mathbf{K}}+i\eta}+\cdots\right\}\right]\frac{1}{\Omega
-E_{\nu,\mathbf{p},\mathbf{K}}+i\eta}\ ,\hspace{0.5cm}
\end{eqnarray}
can be formally rewritten as
\begin{equation}
A_{\nu'\mathbf{p'}\nu\mathbf{p}}(\Omega,\mathbf{K})=
\left[\delta_{\nu',
\nu}\,\delta_{\mathbf{p'},\mathbf{p}}+\frac{\tilde{W}_{\nu'
\mathbf{p'}\nu\mathbf{p}}(\Omega,\mathbf{K})}
{\Omega-E_{\nu',\mathbf{p'},\mathbf{K}}+i\eta}\right]
\frac{1}
{\Omega-E_{\nu,\mathbf{p},\mathbf{K}}+i\eta}\ .
\end{equation}
$\tilde{W}_{\nu'
\mathbf{p'}\nu\mathbf{p}}(\Omega,\mathbf{K})$ appears as a
``renormalized exciton-electron interaction''. It verifies the
integral equation
\begin{equation}
\tilde{W}_{\nu'\mathbf{p'}\nu\mathbf{p}}(\Omega,\mathbf{K})=
W_{\nu'\nu}(-\mathbf{p'}+\mathbf{p})+\sum_{\nu_1,\mathbf{p}_1}
\frac{\tilde{W}_{\nu'\mathbf{p'}\nu_1\mathbf{p}_1}(\Omega,
\mathbf{K})\,W_{\nu_1\nu}(-\mathbf{p}_1+\mathbf{p})}
{\Omega-E_{\nu_1,\mathbf{p}_1,\mathbf{K}}+i\eta}\ .
\end{equation}
Its iteration is
shown in fig.\ (1). Before going further, let us note that
\begin{equation}
\frac{1}{\Omega-E_{\nu_1,\mathbf{p}_1,\mathbf{K}}+i\eta}=
\int\frac{id\omega_1}{2\pi}g_e(\Omega+\omega_1,\mathbf{p}_1
+\beta_e
\mathbf{K})\,g_x(-\omega_1;\nu_1,-\mathbf{p}_1+\beta_x\mathbf{K})=
G_{xe}(\Omega;\nu_1,\mathbf{p}_1,\mathbf{K})\ ,
\end{equation}
with $g_e(\omega,\mathbf{k})=(\omega-\hbar^2\mathbf{k}^2/2m_e
+i\eta)^{-1}$ being the usual free electron Green's function for
an empty Fermi sea, while $g_x(\omega;n)=(\omega-E_n+i\eta)^{-1}$
is the free boson-exciton Green's function, as if the
excitons were non-interacting bosons, \emph{i}.\ \emph{e}., if
all the
$\Xi$'s
\emph{and} $\Lambda$'s were set equal to zero. $G_{xe}$ thus
appears as the propagator of an exciton-electron pair. It is
quite similar to the electron-hole pair propagator $G_{eh}$
appearing in exciton diagrams (see for example eq.\ 2.10 of ref.\
[10]).

The simplest way to obtain 
$A_{\nu'\mathbf{p'}\nu\mathbf{p}}(\Omega,\mathbf{K})$ is
to insert the trion closure relation between the two operators
of the l.h.s.\ of eq.\ (27) and to project this equation over 
$\langle v|\mathcal{T}_{\nu'',\mathbf{p''},\mathbf{K};S,S_z}$.
By using eqs.\ (19,16,24), we get
\begin{equation}
\sum_{\eta_i}\frac{2\,\psi_{\nu'',\mathbf{p''}}^{(\eta_i;S)}\,
\left(\psi_{\nu,\mathbf{p}}^{(\eta_i;S)}\right)^\ast}
{\Omega-\mathcal{E}_{\eta_i,\mathbf{K};S}+i\eta}=
A_{\nu''\mathbf{p''}\nu\mathbf{p}}(\Omega,\mathbf{K})
+(-1)^S\sum_{\nu',\mathbf{p'}}
L_{\nu''\mathbf{p''}\nu'\mathbf{p'}}\,A_{\nu'\mathbf{p'}\nu
\mathbf{p}}(\Omega,\mathbf{K})\ .
\end{equation}
$A_{\nu''\mathbf{p''}\nu\mathbf{p}}(\Omega,\mathbf{K})$ is then
obtained by adding the above equations for
$S=0$ and
$S=1$. Using eq.\ (29), the renormalized
exciton-electron interaction, \emph{i}.\ \emph{e}., the sum of
the exciton-electron ladder ``rungs'', is thus given by
\begin{equation}
\tilde{W}_{\nu'\mathbf{p'}\nu\mathbf{p}}(\Omega,\mathbf{K})=
\left[-\delta_{\nu',\nu}\,\delta_{\mathbf{p'},\mathbf{p}}+
\frac{1}{G_{xe}(\Omega;\nu',\mathbf{p'},\mathbf{K})}
\sum_{\eta_i,S}\frac{\psi_{\nu',\mathbf{p'}}^{(\eta_i;S)}
\left(\psi_{\nu,\mathbf{p}}^{(\eta_i;S)}\right)^\ast}
{\Omega-\mathcal{E}_{\eta_i,\mathbf{K};S}+i\eta}\right]
\frac{1}{G_{xe}(\Omega;\nu,\mathbf{p},\mathbf{K})}\ .
\end{equation}
This result is quite similar to the ``renormalized electron-hole
Coulomb interaction" appearing in electron-hole ladder diagrams,
as given for example in eq.\ (2.18) of ref.\ (10).

\section{Conclusion}

This work relies on a new
commutation technique for excitons interacting with electrons
which takes exactly into account the possible exchange between
carriers, \emph{i}.\ \emph{e}., the close-to-boson character of
the excitons. Using it, we have generated the trion creation
operator in terms of exciton and electron. We have also
generated the exciton-electron ladder diagram associated to
these trions, in terms of exciton propagators and electron
propagators.

Just as the exciton creation operator in terms of electrons and
holes or the exciton ladder diagram between one electron and
one hole propagators, do not help to solve the
Schr\"{o}dinger equation for one exciton, but allow to deal with
many-body effects involving excitons, the trion creation
operator and the trion ladder diagram derived here are of no use
to solve the Schr\"{o}dinger equation for one trion, but
constitute the appropriate tools for further works on many-body
effects with trions.

\newpage
\hbox to \hsize{\hfill REFERENCES \hfill}
\vspace{1cm}

\noindent
[1] E. HYLLERAAS, Phys.\ Rev.\ \underline{71}, 491 (1947).

\noindent
[2] M. LAMPERT, Phys.\ Rev.\ Lett.\ \underline{1}, 450 (1958).

\noindent
[3] K. KHENG, R.T. COX, Y. MERLE D'AUBIGN\'{E}, F. BASSANI, K.
SAMINADAYAR, S. TATARENKO, Phys.\ Rev.\ Lett.\ \underline{71},
1752 (1993).

\noindent
[4] G. FINKELSTEIN, H.SHTRIKMAN, I. BAR-JOSEPH, Phys.\ Rev.\
Lett.\ \underline{74}, 976 (1995).

\noindent
[5] A.J. SHIELDS, M. PEPPER, D.A. RITCHIE, M.Y. SIMMONS, G.A.
JONES, Phys.\ Rev.\ B \underline{51}, 18049 (1995).

\noindent
[6] S.A. BROWN, J.F. YOUNG, J.A. BRUM, P. HAWRYLAK, Z.
WASILEWSKI, Phys.\ Rev.\ B, Rapid Com. \underline{54}, 11082
(1996).

\noindent
[7] R. KAUR, A.J. SHIELDS, J.L. OSBORNE, M.Y. SIMMONS, D.A.
RITCHIE, M. PEPPER, Phys.\ Stat.\ Sol.\ \underline{178}, 465
(2000).

\noindent
[8] V. HUARD, R.T. COX, K. SAMINADAYAR, A. ARNOULT, S. TATARENKO,
Phys.\ Rev.\ Lett.\ \underline{84}, 187 (2000).

\noindent
[9] M. COMBESCOT, O. BETBEDER-MATIBET, Europhys.\ Lett.\ 
\underline{58}, 87 (2002) ; O. BETBEDER-MATIBET, M. COMBESCOT,
Eur.\ Phys.\ J.\ B \underline{27}, 505 (2002).

\noindent
[10] See for example, O. BETBEDER-MATIBET, M. COMBESCOT, Eur.\
Phys.\ J.\ B
\underline{22}, 17 (2001).

\noindent
[11] See for example, F. ARIAS DE SAAVEDRA, E. BUENDIA, F.J.
GALVEZ, A. SARSA, Eur.\ Phys.\ J. D \underline{2}, 181 (1998).

\noindent
[12] M. COMBESCOT, O. BETBEDER-MATIBET, B.
ROULET, Europhys.\ Lett.\ \underline{52}, 717
(2002).

\noindent
[13] M. COMBESCOT, O. BETBEDER-MATIBET, 
Eur.\ Phys.\ J.\ B \underline{31}, 305 (2003).               

\newpage
\hbox to \hsize{\hfill FIGURE CAPTIONS \hfill}
\vspace{1cm}

\noindent
Fig (1)

(a) Direct Coulomb scattering between one ``free'' exciton and
one free electron.

(b) Renormalized free exciton-electron interaction as given by
the integral equation (29). It corresponds to the sums of one,
two,\ldots ladder ``rungs'' between one ``free'' exciton and one
free electron.

\end{document}